\documentclass[prl,a4paper,reprint,superscriptaddress,showpacs]{revtex4-1}

\pdfoutput=1

\usepackage{amsmath}
\usepackage{graphicx}
\newcommand{\Rvec}{\boldsymbol{R}}

\newcommand{\rvec}{\boldsymbol{r}}
\newcommand{\kvec}{\boldsymbol{k}}
\newcommand{\uvec}{\boldsymbol{u}}

\newcommand{\mvec}{\boldsymbol{m}}

\newcommand{\xivec}{\boldsymbol{\xi}}

\begin{document}

\title{Bilayers of Rydberg atoms as a quantum simulator for
  unconventional superconductors.}

\author{J.P. Hague}
\affiliation{The Open University, Walton Hall, Milton Keynes, MK7 6AA, UK}

\author{C. MacCormick}
\affiliation{The Open University, Walton Hall, Milton Keynes, MK7 6AA, UK}

\begin{abstract}
In condensed matter, it is often difficult to untangle the effects of competing interactions, and this is especially problematic for superconductors. Quantum simulators may help: here we show how exploiting the properties of highly excited Rydberg states of cold fermionic atoms in a bilayer lattice can simulate electron-phonon interactions in the presence of strong correlation - a scenario found in many unconventional superconductors. We discuss the core features of the simulator, and use numerics to compare with condensed matter analogues. Finally, we illustrate how to achieve a practical, tunable implementation of the simulation using `painted spot' potentials.
\end{abstract}

 

\date{12th December 2011}

\maketitle

Cold atom quantum simulators offer an important new approach to the
study of correlated electron phenomena without the limitations of
computational or analytical techniques. For example, cold atoms have
recently been used as quantum simulators to investigate models of
condensed matter such as the Hubbard model of strong local Coulomb
repulsion, which is important for the understanding of cuprate
superconductors \cite{hubbard1963a,bloch2008a}. This has led to direct
observation of important phenomena such as the superfluid to Mott
insulator transition \cite{greiner2002a,campbell2006a}.

Besides the cuprates, there are several superconductors with high
transition temperatures, many of which have important roles for
electron-phonon interactions, and repulsion driven correlated electron
phenomena such as antiferromagnetism. Fulleride superconductors of the
family A$_{3}$C$_{60}$ have phonon driven transition temperatures
\cite{hebard1991a} of up to 40K \cite{palstra1995a}, but also exhibit
antiferromagnetism at appropriate dopings and structures
\cite{ganin2010a}. Superconductivity in bismuthates with $T_C>30K$
\cite{cava1988a} is
probably due to strong couplings of localized electrons to the lattice \cite{bischofs2002a}
(as evidenced by a large isotope shift
\cite{batlogg1988a,hinks1988a}). High transition temperatures have
also been achieved in the borocarbides \cite{cava1994a,hossain1994a}
($T_C=23$K) and chloronitrides \cite{yamanaka1998a} ($T_C=25$K). The
more conventional layered MgB$_2$ and graphite intercalation compounds
are also interesting. Even the cuprate superconductors, where
superconductivity is thought by many to be driven by
anti-ferromagnetic fluctuations \cite{anderson1997a}, show isotope
shifts \cite{zhao2001a} and other effects such as kinks
\cite{lanzara2001a} that may be attributed to a non-trivial
interplay between strong correlations and lattice vibrations.

Owing to the importance of electron-phonon interactions in condensed
matter, reliable numerical methods have been sought, but even very
simplified models \cite{holstein1959a} are extremely hard to
simulate. Simulations either have to deal with the potentially infinite number of phonons associated with even a single electron, or retardation effects if the phonons are integrated out. A key
problem in many advanced materials is that electrons are localized to
atomic orbitals and accordingly the Fermi energy is small. This
localization means that dimensionless electron-phonon couplings are
relatively strong and phonon frequencies can be large at around 10\%
of the Fermi energy. Even moderate couplings with intermediate
frequency phonons can lead to consequences that cannot be predicted
with perturbation theory and other analytics. Such couplings can cause
additional difficulties for numerics, that are often most efficient in
the extreme limits of weak or strong coupling. Recent advances in
continuous time quantum Monte Carlo (CTQMC) cluster impurity solvers
for dynamical mean field theories offer a state of the art for numerical
simulations of Hubbard and Holstein models (for a review see
\onlinecite{gull2011a}). Such simulations are currently limited to
$\sim$ 36 site clusters for Hubbard models and 12 site clusters for
Holstein models, limiting the range of spatial fluctuations that can be simulated in 2D to a few sites. 

Quantum simulation of lattice effects has proved difficult to
implement. The existence of quantum simulators with a high degree of
control over the form of interactions has the potential to provide
significant insight into the subtle interplay between electronic
correlation, lattice vibration and phenomena such as superconductivity
and colossal magnetoresistance. We propose an approach
for simulating fermionic Hubbard models extended to include strong
long-range interactions with lattice vibrations, by discussing Rydberg
states of cold atoms in bilayer lattices. Such a simulator is likely
to shed light on a wide range of unconventional superconductors with
transition temperatures greater than 30K, and might resolve aspects of ongoing debates on cuprate mechanisms.

\begin{figure}
\includegraphics[width=80mm]{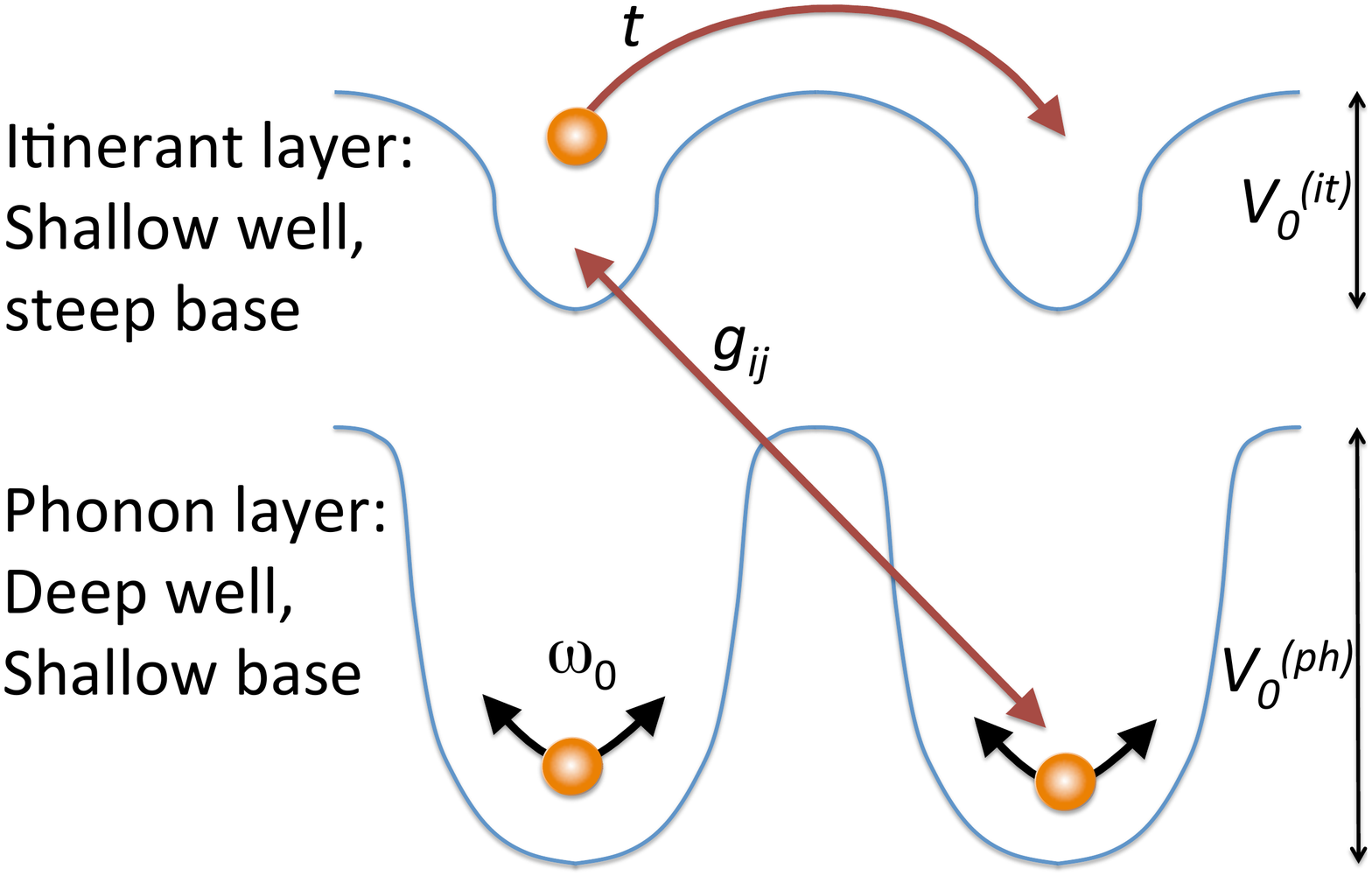}
\caption{System of bilayer Rydberg cold atoms for simulation of strong correlations and interactions between fermions and phonons, annotated with Hamiltonian terms. $t$ is the intersite hopping in the itinerant layer, $\omega_0$ the phonon frequency, $g_{ij}$ the Rydberg-phonon coupling and $V_{0}^{(ph)}$ and $V_{0}^{(it)}$ are trap depths in the itinerant and phonon layers respectively.}
\label{fig:simulator}
\end{figure}

A quantum simulator for complex electronic systems such as
unconventional superconductors should have the following main
characteristics: (1) It must be able to simulate fermions (2) It
should be capable of simulating all filling factors up to and
including half filling (1 fermion per lattice site) where the physics
of materials with complex phase diagrams is most interesting (3) All
parameters, including the phonon frequency, electron-phonon coupling,
Hubbard $U$ and hopping $t$ should be highly tunable for all interesting physical regimes.

Several schemes for quantum simulation of interactions between
electrons and phonons have been suggested. A proposal to bathe bosonic
or fermionic impurities in an optical lattice in a BEC
\cite{bruderer2007a, bruderer2010a, privitera2010a} has led to the
observation of polaron effects for bosonic impurities \cite{gadway2010a}. This succeeds in
criterion (1) as the scheme can in principle treat very low fermion
density. However, it is essential that Fermionic impurities have a negligible effect on the underlying BEC (specifically, fluctuations in the BEC order parameter must be small \cite{bruderer2007a, bruderer2010a, privitera2010a}). This limits the schemes to very low fermion density, voiding criterion (2).

Alternate schemes are based on the interaction between phonons and
bosonic excitations. Interactions with high-energy phonon states of
Rydberg ions are used as part of the mapping in proposed simulators
for spin systems \cite{porras2004a,mueller2008a}. Li and Lesanovsky
have discussed structural distortions associated with exciting high
energy Rydberg states in cold ion crystals \cite{li2011a}. Rydberg
atoms have been proposed as a way of simulating polaron effects in
strongly deformable materials \cite{hague2011a}. The use of cold polar
molecules to obtain Holstein polaron effects has been discussed
\cite{herrera2011a}. These schemes are not extensible to fermions, so
do not fulfil criterion (1), although they are highly tunable. The
scheme proposed in this letter goes beyond these as it is capable of
simulating interacting Fermions with arbitrary filling factor, while
retaining a high level of control over all parameters in the
Hamiltonian, thus satisfying all of critera (1), (2) and (3) necessary
to examine the complex phase diagrams of strongly correlated systems
in the presence of phonons.

We begin by discussing how electron-phonon interactions can be
simulated in a system of cold, highly excited (Rydberg) atoms in a
bilayer lattice. Over long-ranges, we assume that the Rydberg atoms
interact only via dipole-dipole interactions in the strong-coupling
F\"orster regime, $V_{kl} = \mu^2/|\Rvec_{k} -\Rvec_{l}|^3$, where
$\mu$ is the dipole moment on the Rydberg atom and $\Rvec_{k}$ is the
vector to the $k$th atom in the lattice (see supplementary material
for a discussion of the origins of this interaction). The dipole
moment may be written in terms of the coefficient $C_{3}$ as
$\mu=\sqrt{2}C_{3}$, and depends upon the Rydberg states chosen for
the experiment. The ground state atoms have no long range
interaction. By coupling the ground $|g\rangle $ and Rydberg
$|r\rangle $ states with a laser tuned $\Delta$ from the $|g\rangle
\rightarrow |r\rangle$ transition and with coupling strength $\Omega$
(the Rabi frequency) we mix the states $|g\rangle$ and
$|r\rangle$. This technique \cite{Henkel2010} of dressing the atoms
with the laser means that the trapped, ground state atoms acquire the
characteristics of the Rydberg state, but in a controllable
fashion. In particular, the coefficient $C_{3}$ is replaced by an
effective interaction coefficient which we write as
$C'_{3}=(\Omega/2\Delta) C_{3}$. We note that use of a different
regime of dipole-dipole interactions has been proposed for the
simulation of liquid crystalline phases \cite{quintanilla2009a}.

The Rydberg atoms, must now be confined in a bilayer lattice. The
``itinerant'' layer represents the electrons in a condensed
matter problem, and the ``phonon'' layer generates phonon mediated interactions. The itinerant layer can have any filling, and tunneling between adjacent sites is allowed. It is
important that the phonon layer has 1
atom per site and that the tunneling is forbidden - the Mott
insulator phase \cite{greiner2002a}. This is achieved by making the
potential barrier in the itinerant layer smaller than in the phonon
layer. A further complication arises because the atoms in the phonon
layer must be set with low oscillating frequencies, while the
itinerant layer must be set up with high phonon frequencies. This can
be achieved if the optical lattice in the phonon layer has a special
configuration as seen in Fig. \ref{fig:simulator}. This form may be
achieved using painted potentials \cite{henderson2009}.

In the itinerant layer, fermions hop with amplitude $t$, and
experience a local Hubbard $U$, which originates from scattering from
the hard-core potential between the fermions when they share the same
lattice site \cite{bloch2008a}. The Hamitonian for these interactions
is $H_{\rm Hub}=-t\sum_{\langle i,i'
  \rangle}c^{\dagger}_{i}c_{j}+U\sum_{i}n_{i}n_{i}$ ($n_i$ is the
number operator for fermions on site $i$, and $c^{\dagger}_{i}$
creates a Rydberg atom on site $i$). Lattice vibrations are
introduced by displacing the atoms in the phonon layer. Atomic
displacements do not affect the optical lattice, so the vibrations of
the atoms are momentum independent Einstein phonons with Hamiltonian,
%
$H_{\rm ph}=\sum_{\nu}\hbar\omega_{\kvec\nu}(d^{\dagger}_{\nu\kvec}d_{\nu\kvec}+1/2)$
%
and with polarization vectors $\xivec_{\kvec\nu} = \xivec_{\nu}$ in
orthogonal directions. Here $\omega_{\kvec,\nu} = \omega_{\nu}$ is the
angular frequency of a phonon in mode $\nu$ with momentum $\kvec$
and $d^{\dagger}$ and $d$ create and
annihilate phonons.

Small in-plane phonon displacements, $\uvec_i$, cause the interaction
between Rydberg states to become $V'_{kl} =
-\Omega^2\mu^2/4\Delta^2|\Rvec_k+\uvec_k -\Rvec_l-\uvec_l|^3$, which
expands as,
\begin{equation}
V'(\Rvec+\uvec) \approx \frac{\Omega^2\mu^2}{4\Delta^2|\Rvec|^3}-\frac{3\Omega^2\mu^2 \uvec\cdot\hat{\Rvec}}{4\Delta^2|\Rvec|^4}.
\end{equation}

The phonons are quantized by substituting
$\uvec_{i}=\sum_{\kvec,\nu}\sqrt{\hbar/2NM\omega_{\kvec,\nu}}\xivec_{\kvec,\nu}(d_{\kvec,\nu}e^{-i\kvec\cdot\Rvec_i}+d^{\dagger}_{\kvec,\nu}e^{i\kvec\cdot\Rvec_i})$. Since
all sites in the phonon layer are occupied, and the modes momentum
independent, a multimode Rydberg-phonon interaction with extended
Holstein form is derived,
\begin{equation}
H_{\rm R-ph} = \frac{3\mu^2\Omega^2}{4\Delta^2}\left(\frac{\hbar}{2M\omega_0}\right)^{1/2} \sum_{\nu}\sum_{ij} \frac{\hat{\Rvec}_{ij}\cdot \xivec_{\nu}}{R_{ij}^4}
n_{i}(d^{\dagger}_{j,\nu}+d_{j,\nu})
\label{eqn:holsteinext}
\end{equation}
Here, $\Rvec_{ij}$ is a vector between an atom in the itinerant layer
at site $i$, and an atom in the phonon layer at site $j$, $N$ the number of sites and $M$ is the mass of the
atoms. The presence of multiple phonon modes in the Hamiltonian is interesting since such interactions are difficult to simulate
with current numerical techniques.

\begin{figure}
\includegraphics[width=80mm]{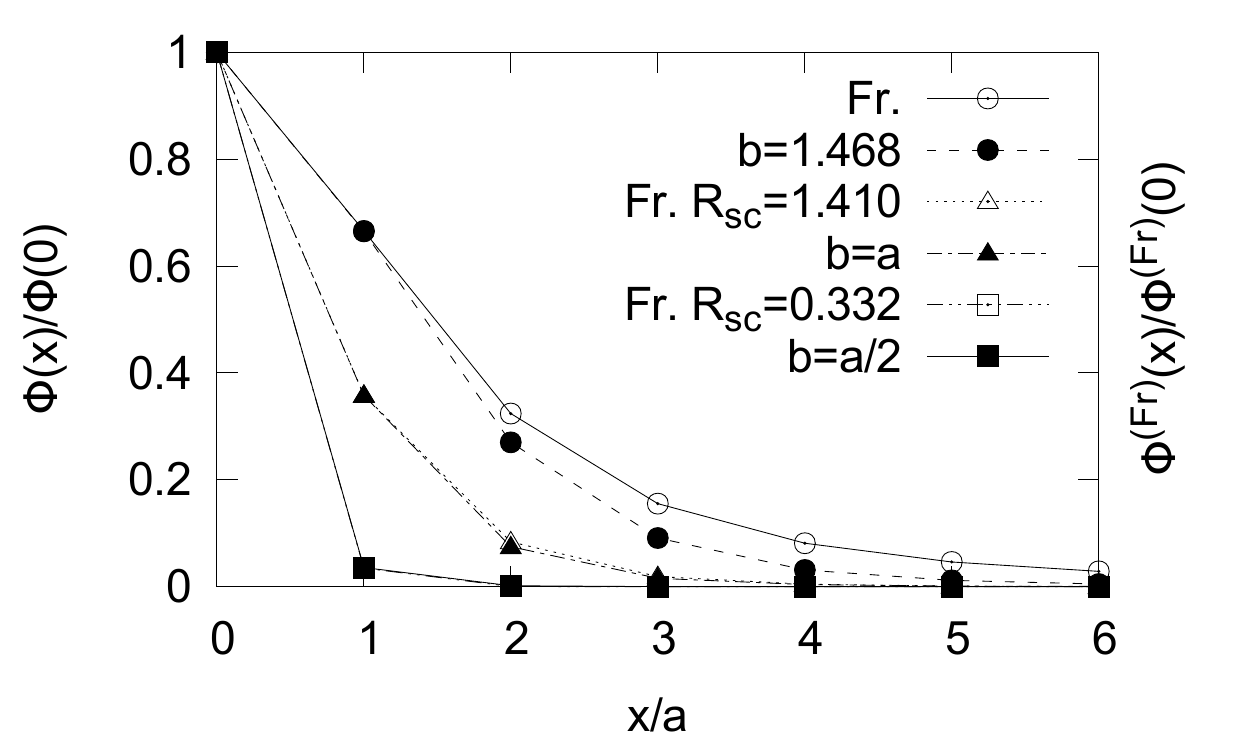}
\caption{Comparison between effective Rydberg-Rydberg interactions in
  the quantum simulator, $\Phi(x)/\Phi(0)$, mediated via phonons
  in the Rydberg atom system here, and the effective electron-electron
  interaction, $\Phi^{\rm(Fr)}(x)/\Phi^{\rm(Fr)}(0)$, mediated by phonons in a typical
  condensed matter system with Fr\"ohlich interactions for various
  interplane distances. An good correspondence between the effective
  interactions is seen.}
\label{fig:effint}
\end{figure}

A further simplification can be made by elongating the potentials in
the phonon layer along the direction perpendicular to the planes, so
that the Hamiltonian becomes,
%
$H_{\rm Holstein} = \sum_{ij} g_{ij} n_{i}(d^{\dagger}_{j}+d_{j})$
%
with the interaction,
$g_{ij} = \frac{\Omega^2}{4\Delta^2} \frac{3\mu^2 b}{(b^2 + r_{ij}^2)^{5/2}}\left(\frac{\hbar}{2M\omega_0}\right)^{1/2}$.
%
where $r_{ij}$ is the distance between the projection of sites $i$ and
$j$ onto the same layer and $b$ is the interplane distance. To
compare with condensed matter analogues, a Holstein model,
where the local electron density couples to local optical phonon modes
\cite{holstein1959a}, has $g^{\rm(Hol)}_{ij}\sim\delta_{ij}$. In other condensed matter systems,
Fr\"ohlich electron-phonon interaction generalized to lattice models (also known
as the extended Holstein interaction)
 has the form $g^{\rm(Fr)}_{ij}\sim
\exp(-r_{ij}/R_{sc})(b^2 + r_{ij}^2)^{-3/2}$ where $R_{sc}$ is a
screening radius\cite{alexandrov1999a}.

When the phonon energy is much larger than hopping, the effective instantaneous interaction between Rydberg atoms in the
itinerant layer mediated though the phonon layer is $-
zt\lambda\sum_{\mvec}\Phi(\rvec)/\Phi(0)$ where
$\Phi(\rvec)/\Phi(0)=\sum_{\mvec}
g_{\rvec,\mvec}g_{0,\mvec}/\sum_{\mvec}g_{0,\mvec}g_{0,\mvec}$. Fig. \ref{fig:effint} shows a comparison between the
shapes of Rydberg and lattice Fr\"ohlich effective interactions
$\Phi^{\rm(Fr)}(x)/\Phi^{\rm(Fr)}(0)$ for various interplane distances and screening radii
($a$ is the intersite distance in the plane). As would be done in the
experiment, the near-neighbor interactions are matched by modifying
$b$ to get the closest possible correspondence to the Fr\"ohlich
interaction that is to be simulated. An excellent correspondence
between the shapes of the interactions is seen for interplane
distance $b\lesssim a$. The origin of the effective interaction is discussed in the supplement.

\begin{figure}
\includegraphics[width=80mm]{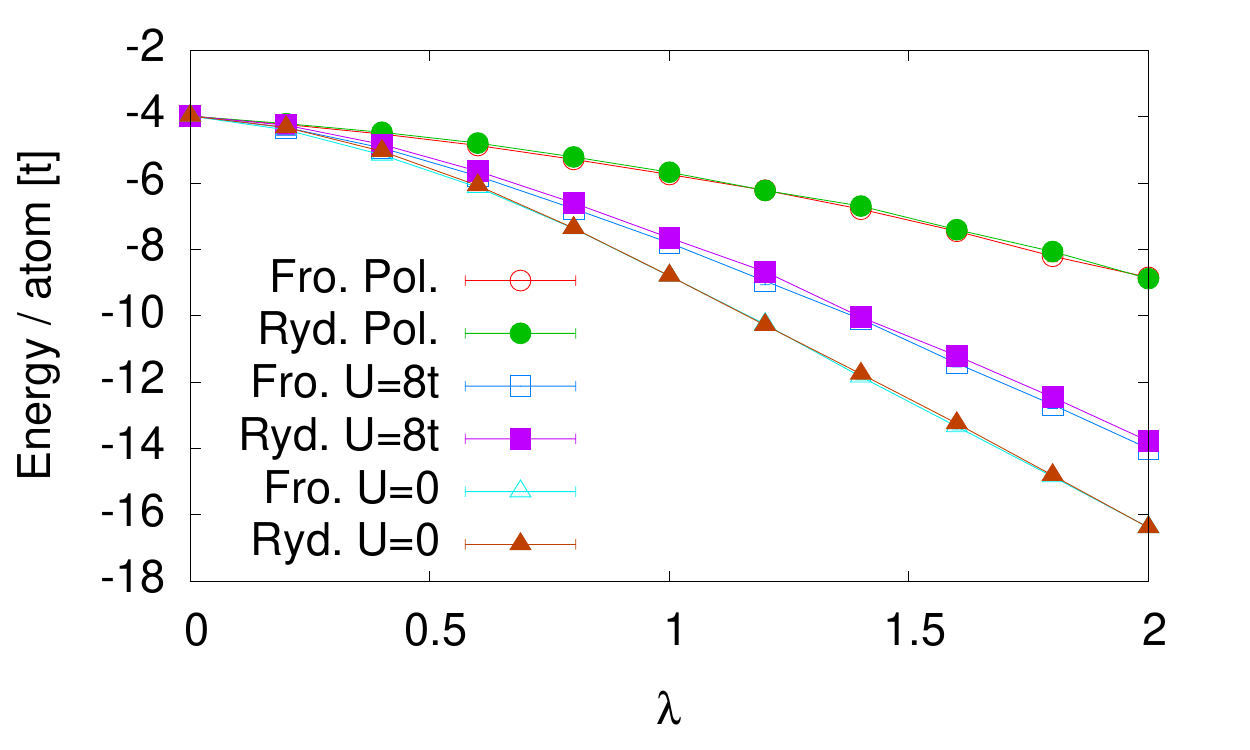}
\caption{Comparison of polaron and bipolaron energies calculated for
  the Rydberg quantum simulator (with $b=1.468a$), and for a screened
  Fr\"ohlich interaction in the condensed matter analogue. $\lambda =
  \Phi(0)/2ztM\omega^2$ and $z$ is the in-plane coordination
  number. To highlight effects of changes in $U$, values of $0$ and
  $8t$ are used. The small differences in the tails of the interaction
  do not strongly affect the physics. $k_B T = 0.014t$, $\omega=0.2t$. Error bars are smaller than
  the points. }
\label{fig:energy}
\end{figure}

To demonstrate the scheme, we use CTQMC \cite{hague2009a} to compute the
properties of polarons and bipolarons in the bilayer lattice and
compare with results from the screened Fr\"ohlich interaction (Fig. \ref{fig:energy}). The simulations include phonon-mediated interactions, the Hubbard
$U$, and also the small direct dipole-dipole interaction between atoms
in the itinerant layer, $H_{\rm direct} = \sum_{i\neq i'}V_{i
  i'}n_{i}n_{i'}$ (as discussed in the supplementary material). A very
close agreement is found between the quantum simulator and Fr\"ohlich
system, further demonstrating that the small differences in the tails
of the interaction and the residual long-range interactions in the itinerant layer do not affect local pairing. 

\begin{figure}
\includegraphics[width=80mm]{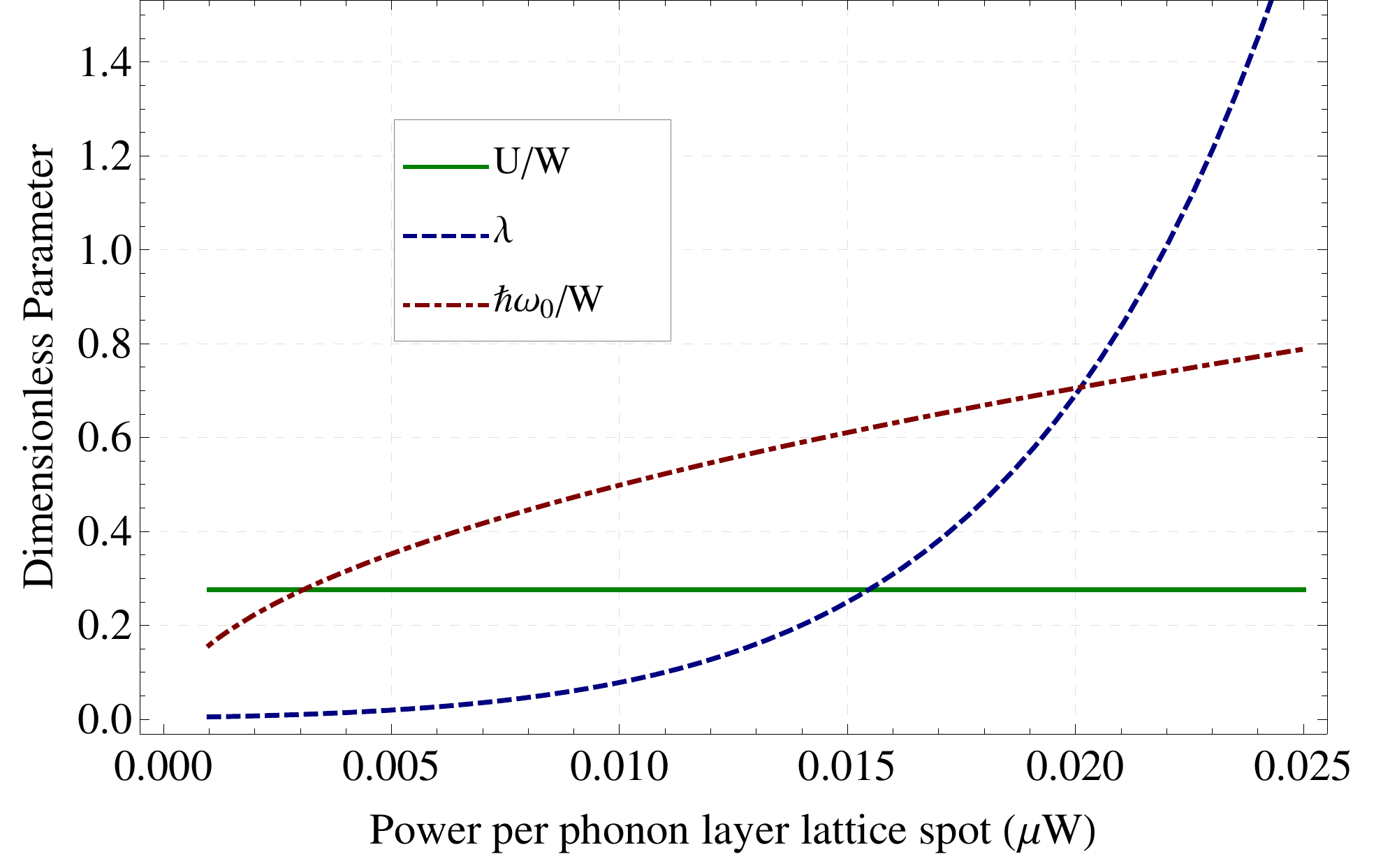}
\caption{Dimensionless parameters for rubidium atoms in the proposed
  spot potential experiment. All physically relevant regimes are
  accessible by tuning the spot power in the itinerant and phonon
  layers ($W=zt$). Further control of $\lambda$ is possible by
  changing the detuning, $\Delta$.}
\label{fig:exptparams}
\end{figure}

Finally, we discuss a realistic experimental setup for building the
quantum simulator. We recommend that experimentalists stage their
investigations by using easier to control bosonic Rydberg atoms before
moving on to fermions. A bilayer optical lattice may be set up with
painted potentials \cite{henderson2009}, a powerful technique with a
high level of control. Creating a bilayer painted lattice requires two
focused horizontal sheets. Gaussian spots of waist $w^{(it)}$ are
focussed on the itinerant sheet, and pairs of spots with waist
$w^{(ph)}$ separated by a distance $2D$ are painted in the phonon
layer. The resulting lattices are filled with Rb atoms and detuned
from the 48S state by $\Delta=125$MHz, with $\Omega=10$MHz leading
to $\mu^2=173$MHz (N.B. The values found here for the bosonic Rb are
representative of all alkali Rydberg atoms). Fig. \ref{fig:exptparams}
shows the dimensionless parameters calculated for a realistic set up,
with $a=2.62\mu$m and $b=3.67\mu$m. Spot lattices in the itinerant
layer have $w^{(it)}=2.46\mu$m, so the potentials overlap leading to a
near sinusoidal potential with $V_{0}^{(it)}=70.1$Hz and energy
level spacing in the itinerant layer of 153Hz, $U=5.34$Hz, $t=4.83$Hz,
so a one band model will be simulated. Phonon
frequencies are easily changed in the phonon layer by tuning $D$, and
can therefore be changed independently. As an example, with
$V_{0}^{(ph)}=578$Hz, $w^{(ph)}=0.655\mu$m and $D=0.999w^{(ph)}$,
$\omega^{(ph)}_0=13.8$Hz and $\lambda=1.00$ is achieved -
challenging but possible with a state of the art diffraction-limited
optical arrangement. A simpler 1D version of the experiment that only
requires a single horizontal sheet can be performed using two coupled
chains (a phonon chain and an itinerant chain) and is recommended as a
starting place when implementing the experiment. In 1D, we find
equally good correspondence between the quantum simulator and
condensed matter analogue.

In the bilayer system, the interesting physics are encoded in the momentum
distribution of the gas held in the itinerant layer. What we want to study
is the correlation function $C(\kvec,\sigma;-\kvec,-\sigma)$. To
extract this, the momentum distribution must be observable. In the
experiments in Ref. \onlinecite{Bakr2010}, this was accomplished using
a time-of-flight method, where the momentum distribution of the
roaming atoms is mapped onto the spatial location at the time of
imaging.

Since it is the purpose of the proposed system to understand phase
diagrams, we briefly note the effect that different interaction
types have in forming and modifying these phases. Repulsive Hubbard $U$ promotes a Mott insulating
state and antiferromagnetism close to half-filling, and may promote
superconductivity through spin fluctuations. It may also control the
form of the superconductivity, reducing or eliminating $s$-wave
pairing. Direct in-plane $V_{ij}$ may promote charge
ordering and suppress superconductivity if it is repulsive, and could enhance superconductivity when attractive. Interaction with phonons
promotes BCS superconductivity at weak coupling, and a BEC of
bipolarons for large $\lambda$.

In this letter, we have shown how systems of cold Rydberg atoms in a
bilayer can be used as a simulator for electron-phonon interactions in
the presence of strong electronic correlation, of the type found in
many unconventional superconductors. We have carried out the mapping
to an extended Hubbard--Holstein model and used numerics to
demonstrate the simulator is capable of reproducing the pairing and
polaron physics of standard electron-phonon models. Furthermore, we
have described how the simulator can be implemented using contemporary
techniques. The proposed system goes well beyond the possibilities of
previous quantum simulators for the simulation of interactions with
lattice vibrations. In particular, we can reliably simulate half
filled femion systems relevant to cuprate and other unconventional
superconductors, tune all parameters directly through the optical
lattice and can easily include multiple phonon modes.


JPH acknowledges EPSRC grant EP/H015655/1 and CM grant
EP/F031130/1. We thank I. Lesanovsky, M. Bruderer, F. Herrera,
A. Kowalczyk, N. Braithwaite, S. Bergamini, S. Alexandrov and
J. Samson for useful discussions, and especially thank Pavel
Kornilovitch for longstanding collaboration on QMC and polarons.

\bibliography{bilayercoldatoms}

\end{document}